\documentclass[10pt,floats,twocolumn,aps,showpacs,preprintnumbers,amsmath,amssymb,prl,superscriptaddress]{revtex4-1}

\usepackage{graphicx}
\usepackage{xcolor}
\usepackage{epsfig}
\usepackage{overpic}
\usepackage{soul}
\usepackage{ulem}
\usepackage[colorlinks]{hyperref} 
\usepackage{cancel}

\begin{document}
 
\title{Suppression of heating in quantum spin clusters under periodic driving as a dynamic localization effect}
\author{Kai Ji}
\email{kji@shnu.edu.cn}
\affiliation{Skolkovo Institute of Science and Technology, Nobel Street 3, 143026 Moscow Region, Russia}
\affiliation{Department of Physics, Shanghai Normal University, No. 100 Guilin Road, 200234 Shanghai, China}
\author{Boris V. Fine}
\email{b.fine@skoltech.ru}
\affiliation{Skolkovo Institute of Science and Technology, Nobel Street 3, 143026 Moscow Region, Russia}
\affiliation{Institut f\"ur Theoretische Physik, Universit\"at Heidelberg, Philosophenweg 19, D-69120 Heidelberg, Germany}

\date{\today}

\begin{abstract}
We investigate numerically and analytically the heating process in ergodic clusters of interacting spins 1/2 subjected to periodic pulses of external magnetic field. Our findings indicate that there is a threshold for the pulse strength below which the heating is suppressed.
This threshold decreases with the increase of the cluster size, approaching zero in the thermodynamic limit; yet it should be observable in clusters with fairly large Hilbert spaces. We obtain the above threshold quantitatively as a condition for the breakdown of the golden rule in the second-order perturbation theory. It is caused by the phenomenon of dynamic localization.

\end{abstract}

\maketitle


Dynamics of many-particle quantum systems under periodic perturbations is a fascinating subject that has recently attracted renewed attention in various contexts. Examples include solid-state nuclear magnetic resonance (NMR) in the presence of continuously applied radio frequency field \cite{Abragam-1961,Kropf-2012}, dynamics of ultra-cold atoms in the presence of oscillating laser potential \cite{Eckardt-2017}, responses of electron-nuclear systems to optical pumping \cite{Petrov-2012}, Floquet topological insulators \cite{Kitagawa-2010,Lindner-2011,Grushin-2014} and laser-driven multiferroics \cite{Sato-2016}.
Periodic driving of qubit systems is also involved in proposals to engineer quantum simulators \cite{Lloyd-1996}. An important fundamental issue in this general context is to identify peculiarly quantum behavior under periodic driving \cite{Moessner-2017}. A relevant quantum phenomenon here is dynamic localization (DL), which is the time-domain analog of Anderson localization \cite{Anderson-1958,Anderson-1978,Lee-1985}.
DL is well understood for systems with one or a few degrees of freedom, such as kicked quantum rotator \cite{Fishman-1982,Grempel-1984,Casati-1987,Buchleitner-1993,Moore-1995,Manai-2015,Chabe-2008,Lopez-2012}.
The current effort is to understand the applicability of DL to many-particle systems.

DL in many-particle systems, if it occurs, would imply that the system stops absorbing energy, which, in turn, would be contrary to the continuous absorption picture based on the standard linear response theory \cite{Landau-1980}.
On the other hand, the absence of the continuous energy absorption is often implied, when the system is subjected to fast and strong periodic driving.
In this case, the often-used framework is the average-Hamiltonian theory, which can be justified by the Magnus expansion \cite{Blanes-2009}, which, in turn, has unclear applicability limits.

A lot of attention in recent years has been focused on the response of systems with strong spatial disorder to periodic driving\cite{Cadez-2017,Luitz-2017,Agarwala-2017}.
Various studies  \cite{Lazarides-2015,Ponte-2015,Khemani-2016,Rehn-2016,Bordia-2017} converged to the conclusion that systems that exhibit many-body localization without periodic driving can also exhibit many-body dynamical localization under periodic driving, when the driving strength is not too large. We, on the other hand, are primarily interested in the periodic driving of the systems, which are ergodic (i.e. thermalizable) without driving. 

If one takes an ergodic isolated many-particle quantum system and starts kicking it, then one would reasonably expect (on the basis of the second law of thermodynamics) that the system absorbs energy after each kick, and, therefore, the temperature of the system gradually increases without any limitation from above.
If the time delay between the kicks is very long, one should further expect that it does not matter whether the kicks are strictly periodic in time or not.
However, it has been shown in our previous numerical investigation, that, surprisingly, the latter expectation does not hold for a system containing 16 spins 1/2 \cite{Ji-2011}.
We found that the heating caused by the periodic kicks with very long delays was asymptotically much slower than that caused by slightly aperiodic kicks.
We attributed the above difference to DL.
The system of 16 spins 1/2 has $2^{16}$ quantum levels, leading one to suspect that the above quantum effects might survive in the thermodynamic limit.
Later studies, however, arrived to the conclusion that weak periodic driving of macroscopic ergodic systems leads to stationary states essentially indistinguishable from the infinite temperature state \cite{DAlessio-2014,Lazarides-2015}.
Such a conclusion is also supported in the present work. Yet, even in such a case, a practically important question remains concerning the size dependence of the DL effects. This question has also been discussed in the literature \cite{Seetharam-2017}, but so far no quantitative criterion for the onset of dynamical localization as a function of the system size and the strength of the perturbation has been formulated.  The present work aims at filling this gap. 
We develop a golden-rule-like theory of heating under the periodic driving and formulate the condition for the break-down of that theory as a function of cluster size and perturbation strength. We verify this criterion by direct numerical simulations.
The above break-down is accompanied by the emergence of quantum corrections that can be detected through the suppression of heating under periodic driving when compared with aperiodic driving.
We propose that the above difference can be used in practice to diagnose the sizes of quantum nanoclusters.

\begin{figure}[t]\setlength{\unitlength}{0.1cm}
\begin{picture} (88,18)
\put(13,-2){\includegraphics[width=0.36\textwidth]{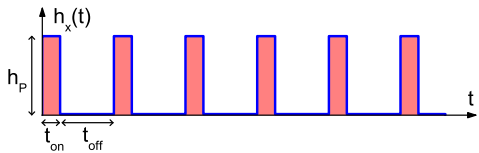}}
\end{picture}
\caption{(Color online) Schematic plot of magnetic field pulses.}
\label{fig:Pulse}
\end{figure}

We consider a quantum spin-1/2 XXZ chain subjected to periodic pulses of external magnetic field.  The system is governed by Hamiltonian
\begin{eqnarray}
\label{eq:mod1}
{\cal H}(t) & = & \sum_{i=1}^{N_s} \left( J_x S_i^x S_{i+1}^x + J_y S_i^y S_{i+1}^y + J_z S_i^z S_{i+1}^z \right)
  \nonumber \\
  & & + \sum_{i=1}^{N_s} h_i^z S_i^z + h_x(t) \sum_{i=1}^{N_s} S_i^x ,
\end{eqnarray}
where  $N_s$ is the total number of spins, $S_i^{\alpha}$($\alpha=x,y,z$) are the spin operators on the $i$th lattice site, and $J_{\alpha}$ are the nearest neighbor coupling constants with values \mbox{$J_x$=$J_y$=-1} and $J_z$=2 chosen such to mimic the NMR experiments in solids \cite{Abragam-1961}, $h_i^z$ are the small static magnetic fields randomly chosen from interval $(-0.2, 0.2)$, and, finally, $h_x(t)$ is the external magnetic field along the $x$-direction, which is  switched on with large amplitude $h_{\rm P} = 10$ for very short time $t_{\rm on} \sim 0.01$ and then switched off for very long time $t_{\rm off} \sim 100$ --- see the sketch in Fig.~\ref{fig:Pulse}. 
We set $\hbar=k_{\rm B}=1$.

The above setting is such that, during each pulse, the perturbation term dominates the dynamics; yet, due to the very small duration of the pulse, the overall effect of the perturbation is very small ($h_{\rm P} t_{\rm on} \ll 1$). We have used a similar setting in Ref.~\cite{Ji-2011}.  The only difference now is the addition of small disordered local fields $h_i^z$, which break the translational invariance of the Hamiltonian and hence make the entire Hilbert space of $2^{N_s}$ states connected by the perturbation. In Ref.~\cite{Ji-2011} the size of the connected blocks in the Hilbert space was smaller than $2^{N_s}$ by factor of roughly $N_s$, which created conditions more favorable to DL \cite{supplement}.

\begin{figure}[t]
\centering
\includegraphics[width=0.45\textwidth]{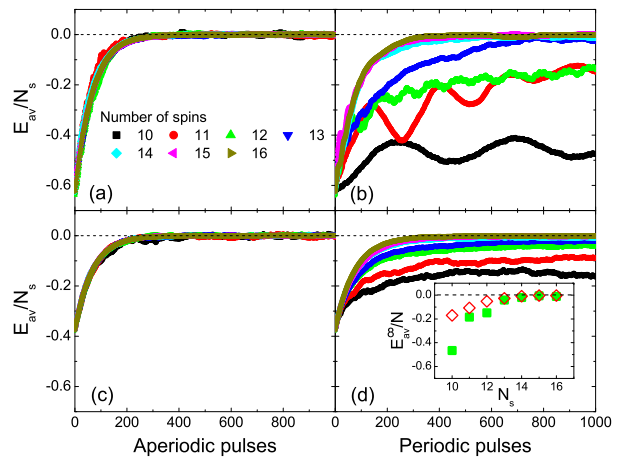}
\caption{(Color online) Average energy per spin for aperiodic (a,c) and periodic (b,d) pulse sequences. The initial state in panels (a,b) is  the ground state; in panels (c,d)  thermal state with $T$=1.
The inset in (d) shows the asymptotic long time values of $E_{\rm av} / N_s$ under periodic driving from the ground state (green squares) and the thermal state (red diamonds).}
\label{fig:Eav}
\end{figure}

Following Ref.~\cite{Ji-2011} let us illustrate the DL effect by comparing the heating process in our system under periodic and aperiodic pulses. We do this by exactly diagonalizing the Hamiltonian (\ref{eq:mod1}) up to 16 spins.
Fig.~\ref{fig:Eav} presents the evolutions of average energy $E_{\rm av}$ per spin for spin chains of different length initially put either in the ground state of Hamiltonian ${\cal H}_{\rm off}$ [panels (a,b)], or in the thermal state \cite{supplement} with temperature $T=1$ [panels (c,d)]. 
Panels (a) and (c)  present respective results for
a series of slightly aperiodic pulses  with $t_{\rm off}$ randomly chosen in the range $100 \pm 10$.
In this case, the heating processes for different $N_s$ all exhibit the same heating per spin towards the infinite temperature limit.
In contrast, significant differences between different $N_s$  arise in panels (b) and (d) representing the behavior under periodic driving.
In particular, we see that the infinite temperature regime associated with  $E_{\rm av} / N_s = 0$ is not reached by clusters of size $N_s=13$ and below. 

In the rest of this paper, we develop a theory of the above DL effect. We only consider periodically pulsed  $h_x(t)$ with period ${\mathcal T} \equiv t_{\rm on}+t_{\rm off} $, where $t_{\rm off}$=100.

Let us denote the initial wave function as $\Psi_0$.
After applying $n$ periodic pulses, the system evolves to a state $\Psi_n = U(\mathcal{T})^n \Psi_0$, where $U ({\mathcal T})$ is the time evolution operator, which is related to a time-independent effective Floquet Hamiltonian ${\cal H}_{\rm eff}$ as 
\begin{eqnarray}
\label{eq:floq2}
U ({\mathcal T}) = e^{-i {\cal H}_{\rm off} t_{\rm off}} e^{-i {\cal H}_{\rm on} t_{\rm on}}
 \equiv e^{-i {\cal H}_{\rm eff} {\mathcal T}} ,
\end{eqnarray}
where ${\cal H}_{\rm on}$  and ${\cal H}_{\rm off}$ correspond to Hamiltonian (\ref{eq:mod1})  with $h_x = h_{\rm P}$ and $h_x= 0$ respectively. We also introduce Floquet phase operator $\Phi \equiv {\cal H}_{\rm eff} {\mathcal T}$, confining its eigenvalues $\phi_\mu$ in the range of $[0,2\pi)$, and then
define the Floquet {\it quasienergies} (eigenvalues of ${\cal H}_{\rm eff}$) as $\{ \varepsilon_{\mu} \equiv \phi_\mu/ {\mathcal T} \}$, which fall into the ``first Floquet zone'' $[0, E_{\rm F}$), where \mbox{$E_{\rm F}\equiv 2 \pi / {\mathcal T}$}.


\begin{figure*}[t] \setlength{\unitlength}{0.1cm}
\begin{picture} (180,34)
{
\put (18,-2){\includegraphics[width=0.8\textwidth]{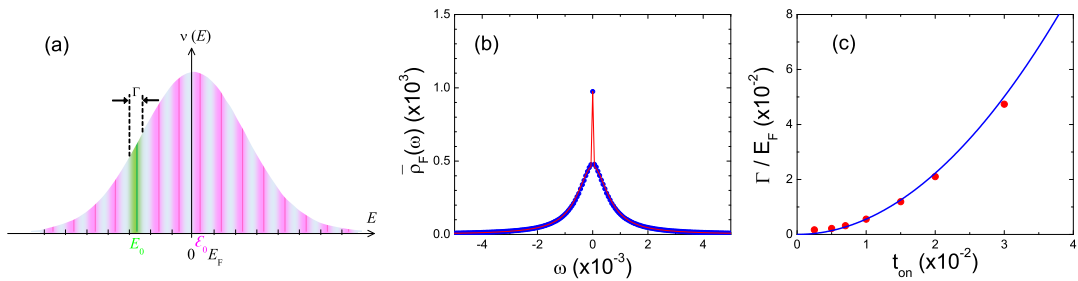}}
}
\end{picture}
\caption{(Color online) (a) Schematic plot of the density of energy states $\nu(E)$ for ${\cal H}_{\rm off}$ with the indication of energy bands of width $\Gamma$  coupled to the initial eigenstate of energy $E_0$ in the process of normal heating by periodic pulses. Adjacent bands are separated by energy $E_F$.   
(b) Average spectral density ${\bar \rho}_{\rm F} (\omega)$ (blue dots) for a chain of 14 spins periodically driven by pulses with $t_{\rm on}$=0.02 and $t_{\rm off}$=100. Red line is the fit by the BW distribution and a $\delta$-function peak at $\omega$=0.
(c) BW width $\Gamma$ for a 14-spin chain with $t_{\rm off}$=100 and different $t_{\rm on}$: red circles --- obtained numerically from ${\bar \rho}_{\rm F} (\omega)$, blue line -  Eq.~(\ref{eq:Gamma}) .
}
\label{fig:BW} 
\end{figure*}

Let us now describe qualitatively the situation of ``normal'' heating.
In this case, an initial eigenstate $\varphi_0$ of the Hamiltonian ${\cal H}_{\rm off}$ with eigenenergy $E_0$ becomes effectively coupled to other eigenstates $\varphi_k$ of ${\cal H}_{\rm off}$ with eigenenergies $E_k$ forming narrow bands around values $E_0 + n E_{\rm F}$, where $n$ is an integer number [see Fig.~\ref{fig:BW}(a)].
The width
of these bands
$\Gamma$ is simultaneously the inverse lifetime of the initial eigenstate $\varphi_0$ under the periodic perturbation, and it also controls the heating rate. In the Floquet theory, one can consider the spectral density $\rho_{\rm F} (\omega) \equiv \sum_{\mu} |\langle \varphi_0 | \psi_{\mu} \rangle|^2 \delta (\omega + \mathcal{E}_0 - \varepsilon_{\mu})$ associated with the decomposition of the initial state  $\varphi_0$ in terms of Floquet eigenstates $\psi_{\mu}$ \cite{Jacquod-1995,Jacquod-1997}.
Here $ \mathcal{E}_0 \equiv E_0 ~{\rm mod}~E_{\rm F}$.
The shape of this spectral density can be well approximated by the Breit-Wigner (BW) formula,
$\rho_{\rm F} (\omega) = (1/ 2 \pi) \Gamma  / (\omega^2 + \Gamma^2 / 4)$.
In Fig.~\ref{fig:BW}(b), we show the above spectral density averaged over all eigenstates of ${\cal H}_{\rm off}$,  $\bar{\rho}_{\rm F}$, for a 14-spin chain.
It can be perfectly fitted by a superposition of a $\delta$-function peak at $\omega=0$ (originating from uninteresting diagonal terms in the perturbation) and the BW distribution.
At the same time, it can be shown analytically from a golden-rule-like calculation \cite{supplement} that, in our setting, the value of $\Gamma$ averaged over all initial eigenstates is
\begin{eqnarray}
\label{eq:Gamma}
\Gamma = \frac{h_{\rm P}^2 t_{\rm on}^2 N_s}{4 \mathcal{T}}.
\end{eqnarray}
In Fig.~\ref{fig:BW}(c), we compare the analytical result (\ref{eq:Gamma}) with the value of $\Gamma$  extracted from the numerically computed $\bar{\rho}_{\rm F}$ for various $t_{\rm on}$ and find a very good agreement.

In the Floquet picture,  the process of heating for the initial wave function $\varphi_0$ is simply the dephasing between different Floquet eigenstates participating in the expansion of $\varphi_0$.
 The participating Floquet eigenstates can, in turn, be decomposed into the eigenstates of the Hamiltonian ${\cal H}_{\rm off}$, forming the energy bands shown in Fig.~\ref{fig:BW}(a), which sample the entire $\nu (E)$. As a result, the participating eigenstates of ${\cal H}_{\rm off}$ fairly represent the infinite-temperature state, even through they constitute only a small fraction $\Gamma / E_{\rm F}$ of all eigenstates of ${\cal H}_{\rm off}$. 

In general, only a minority of the eigenstates of ${\cal H}_{\rm off}$ belonging to the above mentioned energy bands are directly coupled  to $\varphi_0$ by the perturbation ${\cal H}_{\rm P} \equiv h_x(t) \sum_{i=1}^{N_s} S_i^x$. At the same time, only the minority of  the eigenstates of ${\cal H}_{\rm off}$ directly coupled to  $\varphi_0$ would belong to the above energy bands. We formulate the criterion for the normal heating as a self-consistency condition on the validity of the golden-rule formula (\ref{eq:Gamma}) for the BW width. Namely, 
{\it in order for the normal heating to occur, it is necessary that, for a typical eigenstate
$\varphi_0$ of Hamiltonian ${\cal H}_{\rm off}$, there is at least one other eigenstate $\varphi_1$ (with energy $E_1$) directly coupled to $\varphi_0$ by a typical perturbation matrix element, such that 
$ \mathcal{E}_1 \equiv E_1 ~{\rm mod}~E_{\rm F}$ falls within the window $\mathcal{E}_0 \pm \Gamma/2$, where $\Gamma$ is obtained from the golden-rule formula (\ref{eq:Gamma}).}

We then associate the onset of DL with the violation of the above criterion, i.e. 
{\it DL sets in, when the golden-rule-predicted $\Gamma$ is too small, and as a result, none of the the eigenstates of ${\cal H}_{\rm off}$ directly coupled to a typical $\varphi_0$ falls, after backfolding to the first Floquet zone, into the BW  window $\mathcal{E}_0 \pm \Gamma/2$.}

The above DL criterion addresses typical eigenstates of ${\cal H}_{\rm off}$. Therefore, it is applicable to the initial conditions close to the infinite temperature limit, where the density of states of ${\cal H}_{\rm off}$ is the highest and hence the tendency to DL is the weakest. From this perspective, the criterion is, certainly, the sufficient condition for the onset of DL at lower temperatures. It is also, possibly, \mbox{the necessary} condition\cite{supplement}: In finite clusters with ergodic Hamiltonians,  if the analog of our DL criterion is satisfied for low-temperature states $\varphi_0$, but high-temperature states are dynamically delocalized, then  low-temperature states are still likely to ``leak'' to the high-temperature range due to the higher-order effects of the perturbations by ${\cal H}_{\rm on}$ not included in the second-order perturbation theory behind Eq.(\ref{eq:Gamma}). In this scenario, the stronger tendency to DL at lower temperatures only delays the onset of normal heating during the {\it prethermalization} stage \cite{Kuwahara-2016,Else-2017,Weidinger-2017,Abanin-2017a,Abanin-2017b,Zeng-2017,Machado-2017,Seetharam-2017}.

Let us now apply the above criterion to the spin system described by Hamiltonian (\ref{eq:mod1}).
Since both  ${\cal H}_{\rm off}$ and ${\cal H}_{\rm P} $ are  local in the sense that they are sums of local terms,  the perturbation ${\cal H}_{\rm P}$ does not have significant matrix elements coupling the eigenstates of ${\cal H}_{\rm off}$  separated by an energy much larger than a typical one-particle energy of ${\cal H}_{\rm off}$, which can be estimated as the root-mean-squared value of the local field:
$h_{\rm rms} = \left[ N_{NN} \cdot \left( J_x^2 \langle S_x^2 \rangle + J_y^2 \langle S_y^2 \rangle + J_z^2 \langle S_z^2 \rangle \right) \right]^{1/2} = \sqrt{3}$,
where $N_{NN} = 2$ is the number of nearest neighbors for each spin.
If we represent ${\cal H}_{\rm P}$ in the eigenbasis of an ergodic Hamiltonian ${\cal H}_{\rm off}$,  we expect to get a band random matrix which has its non-zero elements located within a diagonal band with a typical half-width of
$h_{\rm rms}$.
For our spin system, the total number of states per unit energy interval, ${\cal N}$, can be approximated as a Gaussian,
$
{\cal N} (E) \approx \left(2^{N_s} / \sqrt{2 \pi} \sigma_E \right) \exp \left(- E^2 / 2 \sigma_E^2\right) 
$,
where
$\sigma_E \approx \left[ N_s \cdot N_{NN} \cdot \left( 
J_x^2 \langle S_{ix}^2 \rangle \langle S_{jx}^2 \rangle + J_y^2 \langle S_{iy}^2 \rangle \langle S_{jy}^2 \rangle 
\right. \right.$ 
$\left. \left. + J_z^2 \langle S_{iz}^2 \rangle \langle S_{jz}^2 \rangle \right) \right]^{1/2} = \sqrt{3 N_s} /2$
is the root-mean-squared value of the total energy.
Thus the relative number of nonzero matrix elements of ${\cal H}_{\rm P}$ in the eigenbasis of ${\cal H}_{\rm off}$  can be estimated as
$h_{\rm rms} / \sigma_E = 2 / \sqrt{N_s}$.

Since the Hamiltonian ${\cal H}_{\rm off}$ is time-independent, it can also be considered as being periodic with period $\mathcal{T}$. The perturbation problem can now be treated in the Floquet quasienergy representation.
After the eigenenergies $E_k$ of ${\cal H}_{\rm off}$ are folded into the first Floquet zone using relation $ \mathcal{E}_k \equiv E_k ~{\rm mod}~E_{\rm F}$, and then the respective eigenstates are ordered according to the value of $\mathcal{E}_k$, the perturbation ${\cal H}_{\rm P}$ becomes a sparse random matrix, where the non-zero elements are uniformly distributed over the entire matrix. 
This is because
$h_{\rm rms} \gg E_F $.
Thus, in each column (or line) of the matrix ${\cal H}_{\rm P}$, the total number of non-zero elements is $N_V \approx 2^{N_s} \cdot 2 / \sqrt{N_s}$.
Thus, in the first Floquet zone, the typical distance between the quasienergies of two unperturbed neighboring states coupled by non-zero elements of ${\cal H}_{\rm P}$  is 
\begin{eqnarray}
\label{eq:crit2}
\Delta {\mathcal E} \approx {E_F \over N_V} = {E_F \sqrt{N_s} \over 2^{N_s} \cdot 2} .
\end{eqnarray}
According to our criterion, normal heating takes place, when,  for a typical unperturbed state $\varphi_0$, the energy window $\mathcal{E}_0 \pm \Gamma/2$ contains at least one other state $\varphi_1$ directly coupled to $\varphi_0$. This implies condition
$\Gamma / \Delta {\mathcal E} \ge 2$, from which, using Eqs.~(\ref{eq:Gamma}) and (\ref{eq:crit2}), we find the threshold value
\begin{eqnarray}
\label{eq:crit4}
t_{\rm on} = {2 \over h_{\rm P}} \sqrt{ 2 \pi \over 2^{N_s} \sqrt{N_s}}.
\end{eqnarray}

\begin{figure}[t]
\centering
\includegraphics[width=0.45\textwidth]{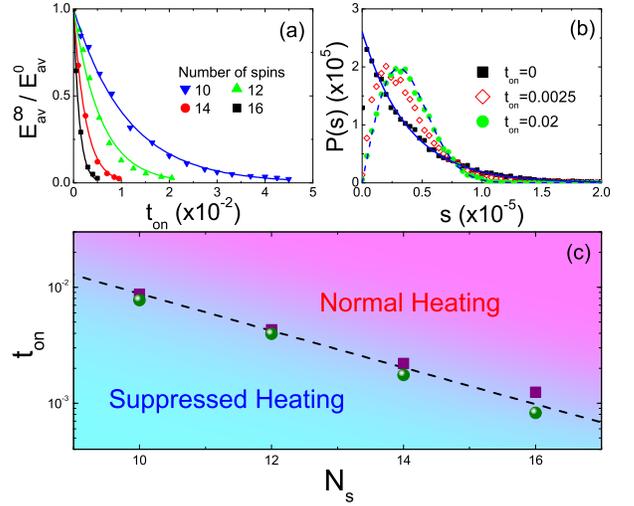}
\caption{(Color online) (a) Ratio $E_{\rm av}^{\infty}/E_{\rm av}^0$ under periodic driving for different $N_s$  as a function of $t_{\rm on}$ averaged over the ensemble of initial states, which are single eigenstates of ${\cal H}_{\rm off}$ \cite{supplement}. Symbols --- numerical results; lines --- exponential fits.  
(b) Probability distributions of spacings $s$ between Floquet quasienergies for a 14-spin chain with different $t_{\rm on}$. Blue solid line --- Poission distribution;  blue dashed line --- Wigner-Dyson distribution.
(c) Phase diagram of the crossover from normal to suppressed heating as a function of the pulse duration $t_{\rm on}$ and the number of spins $N_s$.   Dashed line --- critical condition given  by Eq.~(\ref{eq:crit4}). Symbols are obtained numerically by condition  $E_{\rm av}^{\infty}/E_{\rm av}^0$=0.5. Green balls are obtained from panel (a); purple squares are obtained for the initial thermal state with $T$=1.}
\label{fig:Criterion}
\end{figure}

To test the crossover criterion (\ref{eq:crit4}), we numerically calculate the asymptotic average energy in the long-time limit \cite{supplement}
$E_{\rm av}^{\infty} \equiv \sum_{\mu} |\langle \varphi_0 | \psi_{\mu} \rangle|^2 \langle \psi_{\mu} | {\cal H}_{\rm off} | \psi_{\mu} \rangle$ for different $t_{\rm on}$ and $N_s$ with different  initial wave functions $\varphi_0$. Each $\varphi_0$  represents an  eigenstate of ${\cal H}_{\rm off}$ with energy $E_{\rm av}^0 \equiv \langle \varphi_0| {\cal H}_{\rm off} |\varphi_0 \rangle$.
In Fig.~\ref{fig:Criterion}(a),  the ratio $E_{\rm av}^{\infty} / E_{\rm av}^0$  averaged over a representative ensemble is plotted as a function of $t_{\rm on}$.   The ensemble consists of three statistical samples, each including $\sim1$ percent of $2^{N_s}$ eigenstates of ${\cal H}_{\rm off}$ with initial $E_{\rm av}^0$ selected in the vicinity of three energies ${1 \over 4} E_g$, ${1 \over 2} E_g$, and $E_g$, where $E_g$ is the (negative) energy of the ground state. With such a choice, we avoid large statistical noise near $E_{\rm av}^0 = 0$.
We find that all three samples exhibit nearly the same dependence of $E_{\rm av}^{\infty} / E_{\rm av}^0$ on $t_{\rm on}$ \cite{supplement}, 
which implies that this dependence is representative of a larger ensemble of randomly chosen initial eigenstates of ${\cal H}_{\rm off}$. In turn, the ratio $E_{\rm av}^{\infty} / E_{\rm av}^0$ for the latter characterizes the heating process close to $T= \infty$.

For each $N_s$ in Fig.~\ref{fig:Criterion}(a), the ratio $E_{\rm av}^{\infty} / E_{\rm av}^0$ decreases nearly exponentially with increasing $t_{\rm on}$.
Normal heating corresponds to $E_{\rm av}^{\infty} / E_{\rm av}^0 = 0$, while completely suppressed heating means $E_{\rm av}^{\infty} / E_{\rm av}^0 = 1$.
We then define the value of $t_{\rm on}$ for the crossover between normal heating and DL as the one giving $E_{\rm av}^{\infty} / E_{\rm av}^0 = 0.5$, and plot this value as a function of $N_s$ in Fig.~\ref{fig:Criterion}(c).
The result exhibits a good agreement with the analytical criterion (\ref{eq:crit4}) plotted as a line in Fig.~\ref{fig:Criterion}(c).
Also plotted in Fig.~\ref{fig:Criterion}(c) are the results for the ensemble of initial thermal states with $T=1$ \cite{supplement}.
In accordance with our earlier discussion, we attribute the small difference between the results for $T=1$ and $T= \infty$ to the longer pretermalization stage expected for finite temperatures \cite{supplement}.

An additional piece of evidence that the above suppression of heating  is related to DL  is the statistics of spacings $s$ between adjacent Floquet quasienergies. In  Fig.~\ref{fig:Criterion}(b), we plot this statistics for $N_s$=14 and three values of $t_{\rm on}$  across the crossover from the normal heating to the suppressed heating, and observe the simultaneous crossover from the Wigner-Dyson to the Poisson statistics.

The threshold (\ref{eq:crit4}) for the suppression of heating can be used to determine the number of spins in a nano-cluster.
For a finite cluster, by changing either $t_{\rm on}$ or the perturbation strength, one can observe the onset of significant difference of the heating response of the system to periodic and aperiodic perturbations of equal strengths, and then estimate $N_s$ using Eq.(\ref{eq:crit4}).  Given the exponential dependence on $N_s$ in   Eq.(\ref{eq:crit4}), such an estimate is supposed to be quite accurate.


In conclusion, we formulated the criterion for the onset of quantum suppression of heating in finite clusters, tested this criterion numerically, and illustrated that it is related to the phenomenon of DL.  We further proposed that it can be used to diagnose the size of finite spin clusters.
Our criterion should be generalizable to a broader class of systems and perturbations.  

The authors thank L. D'Alessio, A. Polkovnikov and D. Shepelyansky for enlightening discussions at the early stages of this project.
This work of B.V. F. was supported by a grant of the Russian Science Foundation (Project No. 17-12-01587).
K.J. was partially supported by the Shanghai Pujiang Program (Project No. 17PJ1407400).

\bibliography{heating}

\begin{thebibliography}{44}%
\makeatletter
\providecommand \@ifxundefined [1]{%
 \@ifx{#1\undefined}
}%
\providecommand \@ifnum [1]{%
 \ifnum #1\expandafter \@firstoftwo
 \else \expandafter \@secondoftwo
 \fi
}%
\providecommand \@ifx [1]{%
 \ifx #1\expandafter \@firstoftwo
 \else \expandafter \@secondoftwo
 \fi
}%
\providecommand \natexlab [1]{#1}%
\providecommand \enquote  [1]{``#1''}%
\providecommand \bibnamefont  [1]{#1}%
\providecommand \bibfnamefont [1]{#1}%
\providecommand \citenamefont [1]{#1}%
\providecommand \href@noop [0]{\@secondoftwo}%
\providecommand \href [0]{\begingroup \@sanitize@url \@href}%
\providecommand \@href[1]{\@@startlink{#1}\@@href}%
\providecommand \@@href[1]{\endgroup#1\@@endlink}%
\providecommand \@sanitize@url [0]{\catcode `\\12\catcode `\$12\catcode
  `\&12\catcode `\#12\catcode `\^12\catcode `\_12\catcode `\%12\relax}%
\providecommand \@@startlink[1]{}%
\providecommand \@@endlink[0]{}%
\providecommand \url  [0]{\begingroup\@sanitize@url \@url }%
\providecommand \@url [1]{\endgroup\@href {#1}{\urlprefix }}%
\providecommand \urlprefix  [0]{URL }%
\providecommand \Eprint [0]{\href }%
\providecommand \doibase [0]{http://dx.doi.org/}%
\providecommand \selectlanguage [0]{\@gobble}%
\providecommand \bibinfo  [0]{\@secondoftwo}%
\providecommand \bibfield  [0]{\@secondoftwo}%
\providecommand \translation [1]{[#1]}%
\providecommand \BibitemOpen [0]{}%
\providecommand \bibitemStop [0]{}%
\providecommand \bibitemNoStop [0]{.\EOS\space}%
\providecommand \EOS [0]{\spacefactor3000\relax}%
\providecommand \BibitemShut  [1]{\csname bibitem#1\endcsname}%
\let\auto@bib@innerbib\@empty
\bibitem [{\citenamefont {Abragam}(1961)}]{Abragam-1961}%
  \BibitemOpen
  \bibfield  {author} {\bibinfo {author} {\bibfnamefont {A.}~\bibnamefont
  {Abragam}},\ }\href@noop {} {\emph {\bibinfo {title} {Principles of Nuclear
  Magnetism}}}\ (\bibinfo  {publisher} {Oxford University Press},\ \bibinfo
  {address} {Oxford},\ \bibinfo {year} {1961})\BibitemShut {NoStop}%
\bibitem [{\citenamefont {Kropf}\ and\ \citenamefont
  {Fine}(2012)}]{Kropf-2012}%
  \BibitemOpen
  \bibfield  {author} {\bibinfo {author} {\bibfnamefont {C.~M.}\ \bibnamefont
  {Kropf}}\ and\ \bibinfo {author} {\bibfnamefont {B.~V.}\ \bibnamefont
  {Fine}},\ }\href {\doibase 10.1103/PhysRevB.86.094401} {\bibfield  {journal}
  {\bibinfo  {journal} {Phys. Rev. B}\ }\textbf {\bibinfo {volume} {86}},\
  \bibinfo {pages} {094401} (\bibinfo {year} {2012})}\BibitemShut {NoStop}%
\bibitem [{\citenamefont {Eckardt}(2017)}]{Eckardt-2017}%
  \BibitemOpen
  \bibfield  {author} {\bibinfo {author} {\bibfnamefont {A.}~\bibnamefont
  {Eckardt}},\ }\href {\doibase 10.1103/RevModPhys.89.011004} {\bibfield
  {journal} {\bibinfo  {journal} {Rev. Mod. Phys.}\ }\textbf {\bibinfo {volume}
  {89}},\ \bibinfo {pages} {011004} (\bibinfo {year} {2017})}\BibitemShut
  {NoStop}%
\bibitem [{\citenamefont {Petrov}\ and\ \citenamefont
  {Yakovlev}(2012)}]{Petrov-2012}%
  \BibitemOpen
  \bibfield  {author} {\bibinfo {author} {\bibfnamefont {M.~Y.}\ \bibnamefont
  {Petrov}}\ and\ \bibinfo {author} {\bibfnamefont {S.~V.}\ \bibnamefont
  {Yakovlev}},\ }\href {\doibase 10.1134/S1063776112060131} {\bibfield
  {journal} {\bibinfo  {journal} {J. Exp. Theor. Phys.}\ }\textbf {\bibinfo
  {volume} {115}},\ \bibinfo {pages} {326} (\bibinfo {year}
  {2012})}\BibitemShut {NoStop}%
\bibitem [{\citenamefont {Kitagawa}\ \emph {et~al.}(2010)\citenamefont
  {Kitagawa}, \citenamefont {Berg}, \citenamefont {Rudner},\ and\ \citenamefont
  {Demler}}]{Kitagawa-2010}%
  \BibitemOpen
  \bibfield  {author} {\bibinfo {author} {\bibfnamefont {T.}~\bibnamefont
  {Kitagawa}}, \bibinfo {author} {\bibfnamefont {E.}~\bibnamefont {Berg}},
  \bibinfo {author} {\bibfnamefont {M.}~\bibnamefont {Rudner}}, \ and\ \bibinfo
  {author} {\bibfnamefont {E.}~\bibnamefont {Demler}},\ }\href {\doibase
  10.1103/PhysRevB.82.235114} {\bibfield  {journal} {\bibinfo  {journal} {Phys.
  Rev. B}\ }\textbf {\bibinfo {volume} {82}},\ \bibinfo {pages} {235114}
  (\bibinfo {year} {2010})}\BibitemShut {NoStop}%
\bibitem [{\citenamefont {Lindner}\ \emph {et~al.}(2011)\citenamefont
  {Lindner}, \citenamefont {Refael},\ and\ \citenamefont
  {Galitski}}]{Lindner-2011}%
  \BibitemOpen
  \bibfield  {author} {\bibinfo {author} {\bibfnamefont {N.~H.}\ \bibnamefont
  {Lindner}}, \bibinfo {author} {\bibfnamefont {G.}~\bibnamefont {Refael}}, \
  and\ \bibinfo {author} {\bibfnamefont {V.}~\bibnamefont {Galitski}},\ }\href
  {\doibase 10.1038/nphys1926} {\bibfield  {journal} {\bibinfo  {journal} {Nat.
  Phys.}\ }\textbf {\bibinfo {volume} {7}},\ \bibinfo {pages} {490} (\bibinfo
  {year} {2011})}\BibitemShut {NoStop}%
\bibitem [{\citenamefont {Grushin}\ \emph {et~al.}(2014)\citenamefont
  {Grushin}, \citenamefont {G\'omez-Le\'on},\ and\ \citenamefont
  {Neupert}}]{Grushin-2014}%
  \BibitemOpen
  \bibfield  {author} {\bibinfo {author} {\bibfnamefont {A.~G.}\ \bibnamefont
  {Grushin}}, \bibinfo {author} {\bibfnamefont {A.}~\bibnamefont
  {G\'omez-Le\'on}}, \ and\ \bibinfo {author} {\bibfnamefont {T.}~\bibnamefont
  {Neupert}},\ }\href {\doibase 10.1103/PhysRevLett.112.156801} {\bibfield
  {journal} {\bibinfo  {journal} {Phys. Rev. Lett.}\ }\textbf {\bibinfo
  {volume} {112}},\ \bibinfo {pages} {156801} (\bibinfo {year}
  {2014})}\BibitemShut {NoStop}%
\bibitem [{\citenamefont {Sato}\ \emph {et~al.}(2016)\citenamefont {Sato},
  \citenamefont {Takayoshi},\ and\ \citenamefont {Oka}}]{Sato-2016}%
  \BibitemOpen
  \bibfield  {author} {\bibinfo {author} {\bibfnamefont {M.}~\bibnamefont
  {Sato}}, \bibinfo {author} {\bibfnamefont {S.}~\bibnamefont {Takayoshi}}, \
  and\ \bibinfo {author} {\bibfnamefont {T.}~\bibnamefont {Oka}},\ }\href
  {\doibase 10.1103/PhysRevLett.117.147202} {\bibfield  {journal} {\bibinfo
  {journal} {Phys. Rev. Lett.}\ }\textbf {\bibinfo {volume} {117}},\ \bibinfo
  {pages} {147202} (\bibinfo {year} {2016})}\BibitemShut {NoStop}%
\bibitem [{\citenamefont {Lloyd}(1996)}]{Lloyd-1996}%
  \BibitemOpen
  \bibfield  {author} {\bibinfo {author} {\bibfnamefont {S.}~\bibnamefont
  {Lloyd}},\ }\href {\doibase 10.1126/science.273.5278.1073} {\bibfield
  {journal} {\bibinfo  {journal} {Science}\ }\textbf {\bibinfo {volume}
  {273}},\ \bibinfo {pages} {1073} (\bibinfo {year} {1996})}\BibitemShut
  {NoStop}%
\bibitem [{\citenamefont {Moessner}\ and\ \citenamefont
  {Sondhi}(2017)}]{Moessner-2017}%
  \BibitemOpen
  \bibfield  {author} {\bibinfo {author} {\bibfnamefont {R.}~\bibnamefont
  {Moessner}}\ and\ \bibinfo {author} {\bibfnamefont {S.~L.}\ \bibnamefont
  {Sondhi}},\ }\href {\doibase 10.1038/nphys4106} {\bibfield  {journal}
  {\bibinfo  {journal} {Nat. Phys.}\ }\textbf {\bibinfo {volume} {13}},\
  \bibinfo {pages} {424} (\bibinfo {year} {2017})}\BibitemShut {NoStop}%
\bibitem [{\citenamefont {Anderson}(1958)}]{Anderson-1958}%
  \BibitemOpen
  \bibfield  {author} {\bibinfo {author} {\bibfnamefont {P.~W.}\ \bibnamefont
  {Anderson}},\ }\href {\doibase 10.1103/PhysRev.109.1492} {\bibfield
  {journal} {\bibinfo  {journal} {Phys. Rev.}\ }\textbf {\bibinfo {volume}
  {109}},\ \bibinfo {pages} {1492} (\bibinfo {year} {1958})}\BibitemShut
  {NoStop}%
\bibitem [{\citenamefont {Anderson}(1978)}]{Anderson-1978}%
  \BibitemOpen
  \bibfield  {author} {\bibinfo {author} {\bibfnamefont {P.~W.}\ \bibnamefont
  {Anderson}},\ }\href {\doibase 10.1103/RevModPhys.50.191} {\bibfield
  {journal} {\bibinfo  {journal} {Rev. Mod. Phys.}\ }\textbf {\bibinfo {volume}
  {50}},\ \bibinfo {pages} {191} (\bibinfo {year} {1978})}\BibitemShut
  {NoStop}%
\bibitem [{\citenamefont {Lee}\ and\ \citenamefont
  {Ramakrishnan}(1985)}]{Lee-1985}%
  \BibitemOpen
  \bibfield  {author} {\bibinfo {author} {\bibfnamefont {P.~A.}\ \bibnamefont
  {Lee}}\ and\ \bibinfo {author} {\bibfnamefont {T.~V.}\ \bibnamefont
  {Ramakrishnan}},\ }\href {\doibase 10.1103/RevModPhys.57.287} {\bibfield
  {journal} {\bibinfo  {journal} {Rev. Mod. Phys.}\ }\textbf {\bibinfo {volume}
  {57}},\ \bibinfo {pages} {287} (\bibinfo {year} {1985})}\BibitemShut
  {NoStop}%
\bibitem [{\citenamefont {Fishman}\ \emph {et~al.}(1982)\citenamefont
  {Fishman}, \citenamefont {Grempel},\ and\ \citenamefont
  {Prange}}]{Fishman-1982}%
  \BibitemOpen
  \bibfield  {author} {\bibinfo {author} {\bibfnamefont {S.}~\bibnamefont
  {Fishman}}, \bibinfo {author} {\bibfnamefont {D.~R.}\ \bibnamefont
  {Grempel}}, \ and\ \bibinfo {author} {\bibfnamefont {R.~E.}\ \bibnamefont
  {Prange}},\ }\href {\doibase 10.1103/PhysRevLett.49.509} {\bibfield
  {journal} {\bibinfo  {journal} {Phys. Rev. Lett.}\ }\textbf {\bibinfo
  {volume} {49}},\ \bibinfo {pages} {509} (\bibinfo {year} {1982})}\BibitemShut
  {NoStop}%
\bibitem [{\citenamefont {Grempel}\ \emph {et~al.}(1984)\citenamefont
  {Grempel}, \citenamefont {Prange},\ and\ \citenamefont
  {Fishman}}]{Grempel-1984}%
  \BibitemOpen
  \bibfield  {author} {\bibinfo {author} {\bibfnamefont {D.~R.}\ \bibnamefont
  {Grempel}}, \bibinfo {author} {\bibfnamefont {R.~E.}\ \bibnamefont {Prange}},
  \ and\ \bibinfo {author} {\bibfnamefont {S.}~\bibnamefont {Fishman}},\ }\href
  {\doibase 10.1103/PhysRevA.29.1639} {\bibfield  {journal} {\bibinfo
  {journal} {Phys. Rev. A}\ }\textbf {\bibinfo {volume} {29}},\ \bibinfo
  {pages} {1639} (\bibinfo {year} {1984})}\BibitemShut {NoStop}%
\bibitem [{\citenamefont {Casati}\ \emph {et~al.}(1987)\citenamefont {Casati},
  \citenamefont {Chirikov}, \citenamefont {Shepelyansky},\ and\ \citenamefont
  {Guarneri}}]{Casati-1987}%
  \BibitemOpen
  \bibfield  {author} {\bibinfo {author} {\bibfnamefont {G.}~\bibnamefont
  {Casati}}, \bibinfo {author} {\bibfnamefont {B.~V.}\ \bibnamefont
  {Chirikov}}, \bibinfo {author} {\bibfnamefont {D.~L.}\ \bibnamefont
  {Shepelyansky}}, \ and\ \bibinfo {author} {\bibfnamefont {I.}~\bibnamefont
  {Guarneri}},\ }\href {\doibase 10.1016/0370-1573(87)90009-3} {\bibfield
  {journal} {\bibinfo  {journal} {Phys. Rep.}\ }\textbf {\bibinfo {volume}
  {154}},\ \bibinfo {pages} {77} (\bibinfo {year} {1987})}\BibitemShut
  {NoStop}%
\bibitem [{\citenamefont {Buchleitner}\ and\ \citenamefont
  {Delande}(1993)}]{Buchleitner-1993}%
  \BibitemOpen
  \bibfield  {author} {\bibinfo {author} {\bibfnamefont {A.}~\bibnamefont
  {Buchleitner}}\ and\ \bibinfo {author} {\bibfnamefont {D.}~\bibnamefont
  {Delande}},\ }\href {\doibase 10.1103/PhysRevLett.70.33} {\bibfield
  {journal} {\bibinfo  {journal} {Phys. Rev. Lett.}\ }\textbf {\bibinfo
  {volume} {70}},\ \bibinfo {pages} {33} (\bibinfo {year} {1993})}\BibitemShut
  {NoStop}%
\bibitem [{\citenamefont {Moore}\ \emph {et~al.}(1995)\citenamefont {Moore},
  \citenamefont {Robinson}, \citenamefont {Bharucha}, \citenamefont
  {Sundaram},\ and\ \citenamefont {Raizen}}]{Moore-1995}%
  \BibitemOpen
  \bibfield  {author} {\bibinfo {author} {\bibfnamefont {F.~L.}\ \bibnamefont
  {Moore}}, \bibinfo {author} {\bibfnamefont {J.~C.}\ \bibnamefont {Robinson}},
  \bibinfo {author} {\bibfnamefont {C.~F.}\ \bibnamefont {Bharucha}}, \bibinfo
  {author} {\bibfnamefont {B.}~\bibnamefont {Sundaram}}, \ and\ \bibinfo
  {author} {\bibfnamefont {M.~G.}\ \bibnamefont {Raizen}},\ }\href {\doibase
  10.1103/PhysRevLett.75.4598} {\bibfield  {journal} {\bibinfo  {journal}
  {Phys. Rev. Lett.}\ }\textbf {\bibinfo {volume} {75}},\ \bibinfo {pages}
  {4598} (\bibinfo {year} {1995})}\BibitemShut {NoStop}%
\bibitem [{\citenamefont {Manai}\ \emph {et~al.}(2015)\citenamefont {Manai},
  \citenamefont {Cl\'ement}, \citenamefont {Chicireanu}, \citenamefont
  {Hainaut}, \citenamefont {Garreau}, \citenamefont {Szriftgiser},\ and\
  \citenamefont {Delande}}]{Manai-2015}%
  \BibitemOpen
  \bibfield  {author} {\bibinfo {author} {\bibfnamefont {I.}~\bibnamefont
  {Manai}}, \bibinfo {author} {\bibfnamefont {J.-F. m.~c.}\ \bibnamefont
  {Cl\'ement}}, \bibinfo {author} {\bibfnamefont {R.}~\bibnamefont
  {Chicireanu}}, \bibinfo {author} {\bibfnamefont {C.}~\bibnamefont {Hainaut}},
  \bibinfo {author} {\bibfnamefont {J.~C.}\ \bibnamefont {Garreau}}, \bibinfo
  {author} {\bibfnamefont {P.}~\bibnamefont {Szriftgiser}}, \ and\ \bibinfo
  {author} {\bibfnamefont {D.}~\bibnamefont {Delande}},\ }\href {\doibase
  10.1103/PhysRevLett.115.240603} {\bibfield  {journal} {\bibinfo  {journal}
  {Phys. Rev. Lett.}\ }\textbf {\bibinfo {volume} {115}},\ \bibinfo {pages}
  {240603} (\bibinfo {year} {2015})}\BibitemShut {NoStop}%
\bibitem [{\citenamefont {Chab\'e}\ \emph {et~al.}(2008)\citenamefont
  {Chab\'e}, \citenamefont {Lemari\'e}, \citenamefont {Gr\'emaud},
  \citenamefont {Delande}, \citenamefont {Szriftgiser},\ and\ \citenamefont
  {Garreau}}]{Chabe-2008}%
  \BibitemOpen
  \bibfield  {author} {\bibinfo {author} {\bibfnamefont {J.}~\bibnamefont
  {Chab\'e}}, \bibinfo {author} {\bibfnamefont {G.}~\bibnamefont {Lemari\'e}},
  \bibinfo {author} {\bibfnamefont {B.}~\bibnamefont {Gr\'emaud}}, \bibinfo
  {author} {\bibfnamefont {D.}~\bibnamefont {Delande}}, \bibinfo {author}
  {\bibfnamefont {P.}~\bibnamefont {Szriftgiser}}, \ and\ \bibinfo {author}
  {\bibfnamefont {J.~C.}\ \bibnamefont {Garreau}},\ }\href {\doibase
  10.1103/PhysRevLett.101.255702} {\bibfield  {journal} {\bibinfo  {journal}
  {Phys. Rev. Lett.}\ }\textbf {\bibinfo {volume} {101}},\ \bibinfo {pages}
  {255702} (\bibinfo {year} {2008})}\BibitemShut {NoStop}%
\bibitem [{\citenamefont {Lopez}\ \emph {et~al.}(2012)\citenamefont {Lopez},
  \citenamefont {Cl\'ement}, \citenamefont {Szriftgiser}, \citenamefont
  {Garreau},\ and\ \citenamefont {Delande}}]{Lopez-2012}%
  \BibitemOpen
  \bibfield  {author} {\bibinfo {author} {\bibfnamefont {M.}~\bibnamefont
  {Lopez}}, \bibinfo {author} {\bibfnamefont {J.-F. m.~c.}\ \bibnamefont
  {Cl\'ement}}, \bibinfo {author} {\bibfnamefont {P.}~\bibnamefont
  {Szriftgiser}}, \bibinfo {author} {\bibfnamefont {J.~C.}\ \bibnamefont
  {Garreau}}, \ and\ \bibinfo {author} {\bibfnamefont {D.}~\bibnamefont
  {Delande}},\ }\href {\doibase 10.1103/PhysRevLett.108.095701} {\bibfield
  {journal} {\bibinfo  {journal} {Phys. Rev. Lett.}\ }\textbf {\bibinfo
  {volume} {108}},\ \bibinfo {pages} {095701} (\bibinfo {year}
  {2012})}\BibitemShut {NoStop}%
\bibitem [{\citenamefont {Landau}\ and\ \citenamefont
  {Lifshitz}(1980)}]{Landau-1980}%
  \BibitemOpen
  \bibfield  {author} {\bibinfo {author} {\bibfnamefont {L.~D.}\ \bibnamefont
  {Landau}}\ and\ \bibinfo {author} {\bibfnamefont {E.~M.}\ \bibnamefont
  {Lifshitz}},\ }\href@noop {} {\emph {\bibinfo {title} {Statistical Physics:
  3d edition, part 1}}}\ (\bibinfo  {publisher} {Elsevier
  Butterworth-Heinemann},\ \bibinfo {address} {Burlington},\ \bibinfo {year}
  {1980})\BibitemShut {NoStop}%
\bibitem [{\citenamefont {Blanes}\ \emph {et~al.}(2009)\citenamefont {Blanes},
  \citenamefont {Casas}, \citenamefont {Oteo},\ and\ \citenamefont
  {Ros}}]{Blanes-2009}%
  \BibitemOpen
  \bibfield  {author} {\bibinfo {author} {\bibfnamefont {S.}~\bibnamefont
  {Blanes}}, \bibinfo {author} {\bibfnamefont {F.}~\bibnamefont {Casas}},
  \bibinfo {author} {\bibfnamefont {J.}~\bibnamefont {Oteo}}, \ and\ \bibinfo
  {author} {\bibfnamefont {J.}~\bibnamefont {Ros}},\ }\href {\doibase
  http://dx.doi.org/10.1016/j.physrep.2008.11.001} {\bibfield  {journal}
  {\bibinfo  {journal} {Phys. Rep.}\ }\textbf {\bibinfo {volume} {470}},\
  \bibinfo {pages} {151} (\bibinfo {year} {2009})}\BibitemShut {NoStop}%
\bibitem [{\citenamefont {\v{C}ade\v{z}}\ \emph {et~al.}(2017)\citenamefont
  {\v{C}ade\v{z}}, \citenamefont {Mondaini},\ and\ \citenamefont
  {Sacramento}}]{Cadez-2017}%
  \BibitemOpen
  \bibfield  {author} {\bibinfo {author} {\bibfnamefont {T.}~\bibnamefont
  {\v{C}ade\v{z}}}, \bibinfo {author} {\bibfnamefont {R.}~\bibnamefont
  {Mondaini}}, \ and\ \bibinfo {author} {\bibfnamefont {P.~D.}\ \bibnamefont
  {Sacramento}},\ }\href {\doibase 10.1103/PhysRevB.96.144301} {\bibfield
  {journal} {\bibinfo  {journal} {Phys. Rev. B}\ }\textbf {\bibinfo {volume}
  {96}},\ \bibinfo {pages} {144301} (\bibinfo {year} {2017})}\BibitemShut
  {NoStop}%
\bibitem [{\citenamefont {Luitz}\ \emph {et~al.}(2017)\citenamefont {Luitz},
  \citenamefont {Lev},\ and\ \citenamefont {Lazarides}}]{Luitz-2017}%
  \BibitemOpen
  \bibfield  {author} {\bibinfo {author} {\bibfnamefont {D.~J.}\ \bibnamefont
  {Luitz}}, \bibinfo {author} {\bibfnamefont {Y.~B.}\ \bibnamefont {Lev}}, \
  and\ \bibinfo {author} {\bibfnamefont {A.}~\bibnamefont {Lazarides}},\ }\href
  {\doibase 10.21468/SciPostPhys.3.4.029} {\bibfield  {journal} {\bibinfo
  {journal} {SciPost Phys.}\ }\textbf {\bibinfo {volume} {3}},\ \bibinfo
  {pages} {029} (\bibinfo {year} {2017})}\BibitemShut {NoStop}%
\bibitem [{\citenamefont {Agarwala}\ and\ \citenamefont
  {Sen}(2017)}]{Agarwala-2017}%
  \BibitemOpen
  \bibfield  {author} {\bibinfo {author} {\bibfnamefont {A.}~\bibnamefont
  {Agarwala}}\ and\ \bibinfo {author} {\bibfnamefont {D.}~\bibnamefont {Sen}},\
  }\href {\doibase 10.1103/PhysRevB.95.014305} {\bibfield  {journal} {\bibinfo
  {journal} {Phys. Rev. B}\ }\textbf {\bibinfo {volume} {95}},\ \bibinfo
  {pages} {014305} (\bibinfo {year} {2017})}\BibitemShut {NoStop}%
\bibitem [{\citenamefont {Lazarides}\ \emph {et~al.}(2015)\citenamefont
  {Lazarides}, \citenamefont {Das},\ and\ \citenamefont
  {Moessner}}]{Lazarides-2015}%
  \BibitemOpen
  \bibfield  {author} {\bibinfo {author} {\bibfnamefont {A.}~\bibnamefont
  {Lazarides}}, \bibinfo {author} {\bibfnamefont {A.}~\bibnamefont {Das}}, \
  and\ \bibinfo {author} {\bibfnamefont {R.}~\bibnamefont {Moessner}},\ }\href
  {\doibase 10.1103/PhysRevLett.115.030402} {\bibfield  {journal} {\bibinfo
  {journal} {Phys. Rev. Lett.}\ }\textbf {\bibinfo {volume} {115}},\ \bibinfo
  {pages} {030402} (\bibinfo {year} {2015})}\BibitemShut {NoStop}%
\bibitem [{\citenamefont {Ponte}\ \emph {et~al.}(2015)\citenamefont {Ponte},
  \citenamefont {Chandran}, \citenamefont {Papić},\ and\ \citenamefont
  {Abanin}}]{Ponte-2015}%
  \BibitemOpen
  \bibfield  {author} {\bibinfo {author} {\bibfnamefont {P.}~\bibnamefont
  {Ponte}}, \bibinfo {author} {\bibfnamefont {A.}~\bibnamefont {Chandran}},
  \bibinfo {author} {\bibfnamefont {Z.}~\bibnamefont {Papić}}, \ and\ \bibinfo
  {author} {\bibfnamefont {D.~A.}\ \bibnamefont {Abanin}},\ }\href {\doibase
  https://doi.org/10.1016/j.aop.2014.11.008} {\bibfield  {journal} {\bibinfo
  {journal} {Ann. Phys.}\ }\textbf {\bibinfo {volume} {353}},\ \bibinfo {pages}
  {196} (\bibinfo {year} {2015})}\BibitemShut {NoStop}%
\bibitem [{\citenamefont {Khemani}\ \emph {et~al.}(2016)\citenamefont
  {Khemani}, \citenamefont {Lazarides}, \citenamefont {Moessner},\ and\
  \citenamefont {Sondhi}}]{Khemani-2016}%
  \BibitemOpen
  \bibfield  {author} {\bibinfo {author} {\bibfnamefont {V.}~\bibnamefont
  {Khemani}}, \bibinfo {author} {\bibfnamefont {A.}~\bibnamefont {Lazarides}},
  \bibinfo {author} {\bibfnamefont {R.}~\bibnamefont {Moessner}}, \ and\
  \bibinfo {author} {\bibfnamefont {S.~L.}\ \bibnamefont {Sondhi}},\ }\href
  {\doibase 10.1103/PhysRevLett.116.250401} {\bibfield  {journal} {\bibinfo
  {journal} {Phys. Rev. Lett.}\ }\textbf {\bibinfo {volume} {116}},\ \bibinfo
  {pages} {250401} (\bibinfo {year} {2016})}\BibitemShut {NoStop}%
\bibitem [{\citenamefont {Rehn}\ \emph {et~al.}(2016)\citenamefont {Rehn},
  \citenamefont {Lazarides}, \citenamefont {Pollmann},\ and\ \citenamefont
  {Moessner}}]{Rehn-2016}%
  \BibitemOpen
  \bibfield  {author} {\bibinfo {author} {\bibfnamefont {J.}~\bibnamefont
  {Rehn}}, \bibinfo {author} {\bibfnamefont {A.}~\bibnamefont {Lazarides}},
  \bibinfo {author} {\bibfnamefont {F.}~\bibnamefont {Pollmann}}, \ and\
  \bibinfo {author} {\bibfnamefont {R.}~\bibnamefont {Moessner}},\ }\href
  {\doibase 10.1103/PhysRevB.94.020201} {\bibfield  {journal} {\bibinfo
  {journal} {Phys. Rev. B}\ }\textbf {\bibinfo {volume} {94}},\ \bibinfo
  {pages} {020201} (\bibinfo {year} {2016})}\BibitemShut {NoStop}%
\bibitem [{\citenamefont {Bordia}\ \emph {et~al.}(2017)\citenamefont {Bordia},
  \citenamefont {L\"{u}schen}, \citenamefont {Schneider}, \citenamefont
  {Knap},\ and\ \citenamefont {Bloch}}]{Bordia-2017}%
  \BibitemOpen
  \bibfield  {author} {\bibinfo {author} {\bibfnamefont {P.}~\bibnamefont
  {Bordia}}, \bibinfo {author} {\bibfnamefont {H.}~\bibnamefont {L\"{u}schen}},
  \bibinfo {author} {\bibfnamefont {U.}~\bibnamefont {Schneider}}, \bibinfo
  {author} {\bibfnamefont {M.}~\bibnamefont {Knap}}, \ and\ \bibinfo {author}
  {\bibfnamefont {I.}~\bibnamefont {Bloch}},\ }\href {\doibase
  10.1038/nphys4020} {\bibfield  {journal} {\bibinfo  {journal} {Nat. Phys.}\
  }\textbf {\bibinfo {volume} {13}},\ \bibinfo {pages} {460} (\bibinfo {year}
  {2017})}\BibitemShut {NoStop}%
\bibitem [{\citenamefont {Ji}\ and\ \citenamefont {Fine}(2011)}]{Ji-2011}%
  \BibitemOpen
  \bibfield  {author} {\bibinfo {author} {\bibfnamefont {K.}~\bibnamefont
  {Ji}}\ and\ \bibinfo {author} {\bibfnamefont {B.~V.}\ \bibnamefont {Fine}},\
  }\href {\doibase 10.1103/PhysRevLett.107.050401} {\bibfield  {journal}
  {\bibinfo  {journal} {Phys. Rev. Lett.}\ }\textbf {\bibinfo {volume} {107}},\
  \bibinfo {pages} {050401} (\bibinfo {year} {2011})}\BibitemShut {NoStop}%
\bibitem [{\citenamefont {D'Alessio}\ and\ \citenamefont
  {Rigol}(2014)}]{DAlessio-2014}%
  \BibitemOpen
  \bibfield  {author} {\bibinfo {author} {\bibfnamefont {L.}~\bibnamefont
  {D'Alessio}}\ and\ \bibinfo {author} {\bibfnamefont {M.}~\bibnamefont
  {Rigol}},\ }\href {\doibase 10.1103/PhysRevX.4.041048} {\bibfield  {journal}
  {\bibinfo  {journal} {Phys. Rev. X}\ }\textbf {\bibinfo {volume} {4}},\
  \bibinfo {pages} {041048} (\bibinfo {year} {2014})}\BibitemShut {NoStop}%
\bibitem [{\citenamefont {Seetharam}\ \emph {et~al.}(2018)\citenamefont
  {Seetharam}, \citenamefont {Titum}, \citenamefont {Kolodrubetz},\ and\
  \citenamefont {Refael}}]{Seetharam-2017}%
  \BibitemOpen
  \bibfield  {author} {\bibinfo {author} {\bibfnamefont {K.}~\bibnamefont
  {Seetharam}}, \bibinfo {author} {\bibfnamefont {P.}~\bibnamefont {Titum}},
  \bibinfo {author} {\bibfnamefont {M.}~\bibnamefont {Kolodrubetz}}, \ and\
  \bibinfo {author} {\bibfnamefont {G.}~\bibnamefont {Refael}},\ }\href
  {\doibase 10.1103/PhysRevB.97.014311} {\bibfield  {journal} {\bibinfo
  {journal} {Phys. Rev. B}\ }\textbf {\bibinfo {volume} {97}},\ \bibinfo
  {pages} {014311} (\bibinfo {year} {2018})}\BibitemShut {NoStop}%
\bibitem [{sup()}]{supplement}%
  \BibitemOpen
  \href@noop {} {\bibinfo  {journal} {See Supplemental Material}\ }\BibitemShut
  {NoStop}%
\bibitem [{\citenamefont {Jacquod}\ and\ \citenamefont
  {Shepelyansky}(1995)}]{Jacquod-1995}%
  \BibitemOpen
\bibfield  {journal} {  }\bibfield  {author} {\bibinfo {author} {\bibfnamefont
  {P.}~\bibnamefont {Jacquod}}\ and\ \bibinfo {author} {\bibfnamefont {D.~L.}\
  \bibnamefont {Shepelyansky}},\ }\href {\doibase 10.1103/PhysRevLett.75.3501}
  {\bibfield  {journal} {\bibinfo  {journal} {Phys. Rev. Lett.}\ }\textbf
  {\bibinfo {volume} {75}},\ \bibinfo {pages} {3501} (\bibinfo {year}
  {1995})}\BibitemShut {NoStop}%
\bibitem [{\citenamefont {Jacquod}\ \emph {et~al.}(1997)\citenamefont
  {Jacquod}, \citenamefont {Shepelyansky},\ and\ \citenamefont
  {Sushkov}}]{Jacquod-1997}%
  \BibitemOpen
  \bibfield  {author} {\bibinfo {author} {\bibfnamefont {P.}~\bibnamefont
  {Jacquod}}, \bibinfo {author} {\bibfnamefont {D.~L.}\ \bibnamefont
  {Shepelyansky}}, \ and\ \bibinfo {author} {\bibfnamefont {O.~P.}\
  \bibnamefont {Sushkov}},\ }\href {\doibase 10.1103/PhysRevLett.78.923}
  {\bibfield  {journal} {\bibinfo  {journal} {Phys. Rev. Lett.}\ }\textbf
  {\bibinfo {volume} {78}},\ \bibinfo {pages} {923} (\bibinfo {year}
  {1997})}\BibitemShut {NoStop}%
\bibitem [{\citenamefont {Kuwahara}\ \emph {et~al.}(2016)\citenamefont
  {Kuwahara}, \citenamefont {Mori},\ and\ \citenamefont
  {Saito}}]{Kuwahara-2016}%
  \BibitemOpen
  \bibfield  {author} {\bibinfo {author} {\bibfnamefont {T.}~\bibnamefont
  {Kuwahara}}, \bibinfo {author} {\bibfnamefont {T.}~\bibnamefont {Mori}}, \
  and\ \bibinfo {author} {\bibfnamefont {K.}~\bibnamefont {Saito}},\ }\href
  {\doibase 10.1016/j.aop.2016.01.012} {\bibfield  {journal} {\bibinfo
  {journal} {Ann. Phys.}\ }\textbf {\bibinfo {volume} {367}},\ \bibinfo {pages}
  {96} (\bibinfo {year} {2016})}\BibitemShut {NoStop}%
\bibitem [{\citenamefont {Else}\ \emph {et~al.}(2017)\citenamefont {Else},
  \citenamefont {Bauer},\ and\ \citenamefont {Nayak}}]{Else-2017}%
  \BibitemOpen
  \bibfield  {author} {\bibinfo {author} {\bibfnamefont {D.~V.}\ \bibnamefont
  {Else}}, \bibinfo {author} {\bibfnamefont {B.}~\bibnamefont {Bauer}}, \ and\
  \bibinfo {author} {\bibfnamefont {C.}~\bibnamefont {Nayak}},\ }\href
  {\doibase 10.1103/PhysRevX.7.011026} {\bibfield  {journal} {\bibinfo
  {journal} {Phys. Rev. X}\ }\textbf {\bibinfo {volume} {7}},\ \bibinfo {pages}
  {011026} (\bibinfo {year} {2017})}\BibitemShut {NoStop}%
\bibitem [{\citenamefont {Weidinger}\ and\ \citenamefont
  {Knap}(2017)}]{Weidinger-2017}%
  \BibitemOpen
  \bibfield  {author} {\bibinfo {author} {\bibfnamefont {S.~A.}\ \bibnamefont
  {Weidinger}}\ and\ \bibinfo {author} {\bibfnamefont {M.}~\bibnamefont
  {Knap}},\ }\href {\doibase 10.1038/srep45382} {\bibfield  {journal} {\bibinfo
   {journal} {Scientific Reports}\ }\textbf {\bibinfo {volume} {7}},\ \bibinfo
  {pages} {45382} (\bibinfo {year} {2017})}\BibitemShut {NoStop}%
\bibitem [{\citenamefont {Abanin}\ \emph
  {et~al.}(2017{\natexlab{a}})\citenamefont {Abanin}, \citenamefont {De~Roeck},
  \citenamefont {Ho},\ and\ \citenamefont {Huveneers}}]{Abanin-2017a}%
  \BibitemOpen
  \bibfield  {author} {\bibinfo {author} {\bibfnamefont {D.~A.}\ \bibnamefont
  {Abanin}}, \bibinfo {author} {\bibfnamefont {W.}~\bibnamefont {De~Roeck}},
  \bibinfo {author} {\bibfnamefont {W.~W.}\ \bibnamefont {Ho}}, \ and\ \bibinfo
  {author} {\bibfnamefont {F.}~\bibnamefont {Huveneers}},\ }\href {\doibase
  10.1103/PhysRevB.95.014112} {\bibfield  {journal} {\bibinfo  {journal} {Phys.
  Rev. B}\ }\textbf {\bibinfo {volume} {95}},\ \bibinfo {pages} {014112}
  (\bibinfo {year} {2017}{\natexlab{a}})}\BibitemShut {NoStop}%
\bibitem [{\citenamefont {Abanin}\ \emph
  {et~al.}(2017{\natexlab{b}})\citenamefont {Abanin}, \citenamefont {De~Roeck},
  \citenamefont {Ho},\ and\ \citenamefont {Huveneers}}]{Abanin-2017b}%
  \BibitemOpen
  \bibfield  {author} {\bibinfo {author} {\bibfnamefont {D.}~\bibnamefont
  {Abanin}}, \bibinfo {author} {\bibfnamefont {W.}~\bibnamefont {De~Roeck}},
  \bibinfo {author} {\bibfnamefont {W.~W.}\ \bibnamefont {Ho}}, \ and\ \bibinfo
  {author} {\bibfnamefont {F.}~\bibnamefont {Huveneers}},\ }\href {\doibase
  10.1007/s00220-017-2930-x} {\bibfield  {journal} {\bibinfo  {journal}
  {Commun. Math. Phys.}\ }\textbf {\bibinfo {volume} {354}},\ \bibinfo {pages}
  {809} (\bibinfo {year} {2017}{\natexlab{b}})}\BibitemShut {NoStop}%
\bibitem [{\citenamefont {Zeng}\ and\ \citenamefont {Sheng}(2017)}]{Zeng-2017}%
  \BibitemOpen
  \bibfield  {author} {\bibinfo {author} {\bibfnamefont {T.-S.}\ \bibnamefont
  {Zeng}}\ and\ \bibinfo {author} {\bibfnamefont {D.~N.}\ \bibnamefont
  {Sheng}},\ }\href {\doibase 10.1103/PhysRevB.96.094202} {\bibfield  {journal}
  {\bibinfo  {journal} {Phys. Rev. B}\ }\textbf {\bibinfo {volume} {96}},\
  \bibinfo {pages} {094202} (\bibinfo {year} {2017})}\BibitemShut {NoStop}%
\bibitem [{\citenamefont {Machado}\ \emph {et~al.}(2017)\citenamefont
  {Machado}, \citenamefont {Meyer}, \citenamefont {Else}, \citenamefont
  {Nayak},\ and\ \citenamefont {Yao}}]{Machado-2017}%
  \BibitemOpen
  \bibfield  {author} {\bibinfo {author} {\bibfnamefont {F.}~\bibnamefont
  {Machado}}, \bibinfo {author} {\bibfnamefont {G.~D.}\ \bibnamefont {Meyer}},
  \bibinfo {author} {\bibfnamefont {D.~V.}\ \bibnamefont {Else}}, \bibinfo
  {author} {\bibfnamefont {C.}~\bibnamefont {Nayak}}, \ and\ \bibinfo {author}
  {\bibfnamefont {N.~Y.}\ \bibnamefont {Yao}},\ }\href
  {https://arxiv.org/abs/1708.01620} {\bibfield  {journal} {\bibinfo  {journal}
  {arXiv:1708.01620}\ } (\bibinfo {year} {2017})}\BibitemShut {NoStop}%
\end{thebibliography}%


\clearpage

\setcounter{equation}{0}
\renewcommand{\theequation}{S\arabic{equation}}
\setcounter{figure}{0}
\renewcommand{\thefigure}{S\arabic{figure}}

\begin{center}
{\bf SUPPLEMENTARY MATERIAL}
\end{center}


\vskip 5mm

Note: Reference numbers in the text below are from the reference list of the main article.

This supplementary material presents the technical details of our calculation and the derivations of some key formulas.

\section{A. Pure state representing temperature $T$.}
\label{sec:T}

Simulations presented in Figs.~\ref{fig:Eav}(c), \ref{fig:Eav}(d) and \ref{fig:Criterion}(c) start with an initial state representing temperature $T=1$. This initial state was generated as follows:
\begin{equation}
| \phi_T \rangle = {1 \over \sqrt{Z_0}} \sum_k \exp\left({-E_k \over 2T} - i \alpha_k \right) | \varphi_k \rangle,
\label{eq:phiT}
\end{equation}
where  $\varphi_k$ is the $k$th eigenstate of ${\cal H}_{\rm off}$ with eigenenergy $E_k$, $\alpha_k$ is a random phase chosen from interval $[0, 2\pi)$, and $Z_0 = \sum_k \exp\left({-E_k \over T}\right)$ is the statistical sum. In principle, a faithful representation of a pure thermal state should also include random fluctuations of the absolute values of quantum amplitudes in Eq.~(\ref{eq:phiT}) around $\exp\left({-E_k \over 2T}\right)$, but we checked that these fluctuations were not essential for the purposes of this manuscript.


\section{B. Details about numerical simulations}
\label{sec:Numerics}

Let us express the wave function after $n$ pulses as $\Psi_n = \sum_k a_k^{(n)} \varphi_k$, where $\varphi_k$ is the eigenstate of $H_{\rm off}$ with an eigenenergy $E_k$,  and $a_k^{(n)}$ is the quantum amplitude.
After applying the next pulse, the system evolves to a new state,
$\Psi_{n+1} = e^{-i {\cal H}_{\rm off} t_{\rm off}} e^{-i {\cal H}_{\rm on} t_{\rm on}} \Psi_n = \sum_k a_k^{(n+1)} \varphi_k$.
It is straightforward to show the new coefficient can be written as
\begin{equation}
a_k^{(n+1)} = \sum_{l,r} e^{-i E_l t_{\rm off}} e^{-i \tilde{E}_r t_{\rm on}} M_{kr}^{\dag} M_{rl} a_l^{(n)},
\label{eq:ak}
\end{equation}
where $\tilde{E}_r$ is the eigenenergy of ${\cal H}_{\rm on}$ for an eigenstate $\tilde{\varphi}_r$, and $M_{rl} \equiv \langle \tilde{\varphi}_r | \varphi_l \rangle$ is the matrix element of unitary transformation.
We simulate the time evolution in Eq.~(\ref{eq:ak}) under both periodic and aperiodic pulse sequences with the help of the complete diagonalization of Hamiltonians ${\cal H}_{\rm on}$ and ${\cal H}_{\rm off}$.

\begin{figure}[t]
\centering
\includegraphics[width=0.3\textwidth]{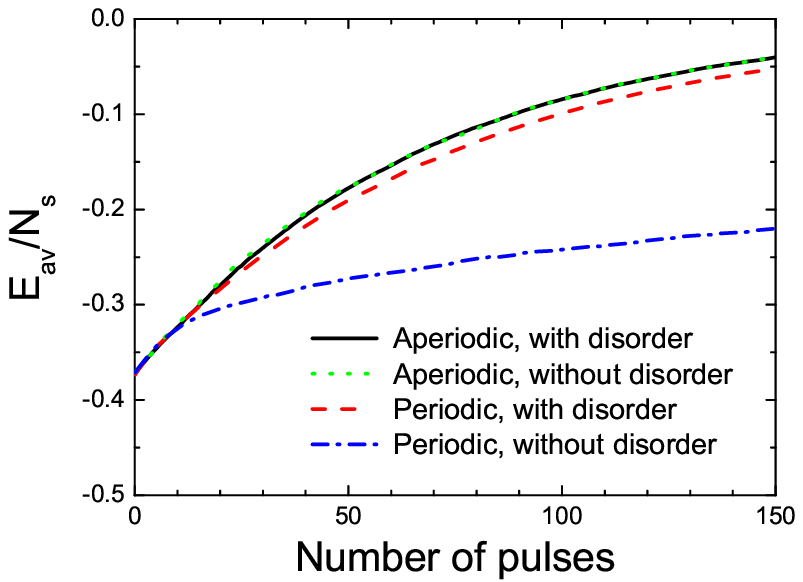}
\caption{
(Color online) Average energy per spin as a function of the number of pulses for a 16-spin cluster with and without  local-field disorder for periodic and aperiodic pulsing.
Black solid and red dashed curves are from Fig.~\ref{fig:Eav} of the main article.
Green dotted and blue dash-dotted curves are from Fig.~4 of Ref.~\cite{Ji-2011}.
}
\label{fig:Disorder}
\end{figure}

In Hamiltonian (\ref{eq:mod1}), we introduced small random local fields $h_i^z$ to break the translational invariance of the system, thereby somewhat suppressing DL and facilitating the normal heating process. We needed this DL suppression, in order to be able to observe the crossover from DL to normal heating for smaller cluster sizes, which are better accessible numerically.
Without the random fields, the quantum basis of the system can be decomposed into $2 N_s$ disconnected subspaces, which reduces the density of dynamically connected quantum states participating in the heating process and hence helps DL.
In Fig.~\ref{fig:Disorder}, we compare the results of $E_{\rm av} / N_s$ shown in Fig.~\ref{fig:Eav} for the disordered 16-spin chain  with those  obtained in Ref.~\cite{Ji-2011} for the same chain without disorder.
In the later case, each of the disconnected subspaces contains only about $2^{11}$ quantum states. Hence the translationally invariant chain exhibits a much stronger DL effect manifested as the larger difference beween the responses to periodic and aperiodic driving.


\section{C. Golden-rule calculation of the heating rate $\Gamma$}
\label{sec:Gamma}

The action of short periodic pulses associated with switching on Hamiltonian 
\begin{eqnarray}
\label{eq:HP}
{\cal H}_{\rm P} = h_{\rm P} \sum_{i=1}^{N_s} S_i^x 
\end{eqnarray} 
can be treated as a small perturbation with respect to the action of Hamiltonian ${\cal H}_{\rm off} $.

In the eigenbasis $\varphi_k$ of ${\cal H}_{\rm off}$, the perturbation induced off-diagonal elements are represented as,
\begin{eqnarray}
\label{eq:Vjk}
V_{jk} = \langle \varphi_j | {\cal H}_{\rm P} | \varphi_k \rangle \, .
\end{eqnarray}

Let us assume that the evolution of wave function starts from the $k$th eigenstate of ${\cal H}_{\rm off}$ with eigenenergy $E_k$, i.e. 
 $\Psi_0 = \varphi_k$. After applying the first pulse, the wave function changes into
\begin{eqnarray}
\label{eq:gr2}
\Psi_1 =C_k^{(1)} \varphi_k + \sum_{j \ne k} C_j^{(1)} \varphi_j,
\end{eqnarray}
where the coefficient $C_j^{(1)}$ can be determined by a perturbative calculation as
\begin{eqnarray}
\label{eq:gr3}
C_j^{(1)} = e^{-i E_j {\mathcal T}}
\left( \delta_{jk} - i t_{\rm on} V_{jk} - {t_{\rm on}^2 \over 2} \sum_l V_{jl} V_{lk} \right) .
\end{eqnarray}
The perturbation due to the pulse slightly depopulates the state $\varphi_k$ reducing its occupation to
\begin{eqnarray}
\label{eq:gr4}
|C_k^{(1)}|^2 = 1 - \sum_{j \ne k} |C_j^{(1)}|^2 \cong 1 - t_{\rm on}^2 \sum_{j \ne k} |V_{jk}|^2 .
\end{eqnarray}
Here the perturbation effect was taken into account up to the second order.
After $n$ short pulses, the occupation of state $\varphi_k$  declines to
\begin{eqnarray}
\label{eq:gr5}
|C_k^{(n)}|^2 \approx \left( 1 - t_{\rm on}^2 \sum_{j \ne k} |V_{jk}|^2 \right)^n \approx e^{- \Gamma_k n {\mathcal T}} ,
\end{eqnarray}
where
\begin{eqnarray}
\label{eq:gr6}
\Gamma_k = {t_{\rm on}^2 \over {\mathcal T}} \sum_{j(j \ne k)} |V_{jk}|^2 .
\end{eqnarray}

The $k$-dependence of $\Gamma_k$ can be eliminated by making an averaging over all states,
\begin{eqnarray}
\label{eq:gr7}
\Gamma \equiv {1 \over N} \sum_k \Gamma_k ,
\end{eqnarray}
where $N$ ($\equiv 2^{N_s}$) is the total number of states.
The constraint `$j \ne k$' in Eq.~(\ref{eq:gr6}) can be neglected, because the number of the diagonal elements of matrix $V_{jk}$ is much fewer than the number of off-diagonal elements, while the values of diagonal elements are comparable with the values of many more off-diagonal elements. [The diagonal elements of $V_{jk}$ are behind the $\delta$-function peak in Fig.~\ref{fig:BW}(b).] This gives us $\sum_k \sum_{j \ne k} |V_{jk}|^2 \approx \sum_{k,j} |V_{jk}|^2 = {\rm Tr} ({\cal H}_{\rm P}^2)$.
The trace can be readily worked out in the eigenbasis of $S_i^x$,
\begin{eqnarray}
\label{eq:gamma6}
{\rm Tr} ({\cal H}_{\rm P}^2) = {h_{\rm P}^2 N_s N \over 4} .
\end{eqnarray}
Substituting it back into Eq.~(\ref{eq:gr7}), we obtain the heating rate as the characteristic width of the Breit-Wigner window
\begin{eqnarray}
\label{eq:gamma7}
\Gamma = \frac{h_{\rm P}^2 t_{\rm on}^2 N_s}{4 {\mathcal T}} \, ,
\end{eqnarray}
which is the same as Eq.~(\ref{eq:crit2}) of the main article.


\section{D. Asymptotic limit of average energy $E_{\rm av}^{\infty}$ under periodic perturbations}
\label{sec:limit}

The long time limit of $E_{\rm av}$, which we denote as $E_{\rm av}^{\infty}$ can be calculated in the following way.
After $n$ pulses, the wave function of the system in the eigenbasis of ${\cal H}_{\rm off}$ can be expressed as
\begin{equation}
\Psi_n = \sum_k C_k^{(n)} \varphi_k .
\label{Psin}
\end{equation}
The corresponding average energy can then be represented as 
\begin{eqnarray}
\label{eq:lim1}
E_{\rm av}^{(n)} & = & \sum_k |C_k^{(n)}|^2 E_k
  = \sum_k |C_k^{(n)}|^2 \langle \varphi_k | {\cal H}_{\rm off} | \varphi_k \rangle
  \nonumber \\
& = & \sum_{\mu, \nu} A_{\nu \mu}^{(n)} \langle \psi_{\mu} | {\cal H}_{\rm off} | \psi_{\nu} \rangle ,
\end{eqnarray}
where $\{ \psi_{\mu} \}$ are the Floquet eigenstates under periodic driving, and
\begin{eqnarray}
\label{eq:lim2}
A_{\nu \mu}^{(n)} = \sum_k \langle \psi_{\nu} | \varphi_k \rangle |C_k^{(n)}|^2 \langle \varphi_k | \psi_{\mu} \rangle .
\end{eqnarray}

The time evolution simply induces a dephasing effect between the Floquet eigenstates, i.e. the off-diagonal matrix elements of $\langle \psi_{\mu} | {\cal H}_{\rm off} | \psi_{\nu} \rangle$ gradually average out as $n$ increases.
Hence, at the limit $n \rightarrow \infty$, we have
\begin{eqnarray}
\label{eq:lim3}
E_{\rm av}^{\infty} = \sum_{\mu} A_{\mu \mu} \langle \psi_{\mu} | {\cal H}_{\rm off} | \psi_{\mu} \rangle .
\end{eqnarray}


\section{E. Ensemble sampling for Fig. 4(a)}
\label{sec:thermal}

\begin{figure}[t]
\centering
\includegraphics[width=0.45\textwidth]{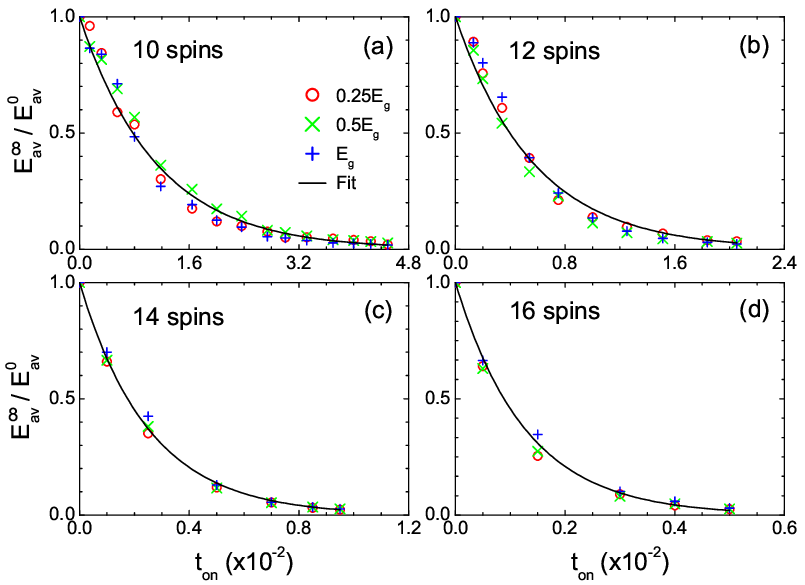}
\caption{
(Color online) Ratio $E^{\infty}_{\rm av} / E^0_{\rm av}$ under periodic driving as a function of $t_{\rm on}$ for different $N_s$.
In each panel, symbols  represent  three different statistical samples with $E^0_{\rm av}$ selected in the vicinity of three energies ${1 \over 4} E_g$,  ${1 \over 2} E_g$, and  $E_g$.
The black lines show the exponential fits to the averages of these three samples.}
\label{fig:Ensemble}
\end{figure}

As mentioned in the main article, the ensemble of initial wave functions used to generate Fig.~\ref{fig:Criterion}(a) 
consists of three different statistical samples, each including  $\sim 1\%$ of $2^{N_s}$ eigenstates of ${\cal H}_{\rm off}$ with $E_{\rm av}^0$ selected in the vicinity of three energies ${1 \over 4} E_g$, ${1 \over 2} E_g$, and
 $E_g$, where $E_g$ is the ground-state energy.
In Fig.~\ref{fig:Ensemble}, we present the statistical average of $E^{\infty}_{\rm av} / E^0_{\rm av}$ separately for each of the above three samples as a function of $t_{\rm on}$ for different $N_s$.  The three samples exhibit nearly the same dependence of $E_{\rm av}^{\infty} / E_{\rm av}^0$ on $t_{\rm on}$. This indicates that the average ensemble used in Fig.~\ref{fig:Criterion}(a) represents the heating response  starting from a typical randomly chosen eigenstate of ${\cal H}_{\rm off}$, which, in turn, implies the initial conditions close to $T= \infty$. 

It may appear, at first sight, surprising, that the sample including the lowest 1 percent of energy eigenstates (the one representing the vicinity of $E_g$) exhibits the heating response representative of a nearly infinite temperature, while the ensemble representing thermal states with $T=1$ exhibits in Fig.\ref{fig:Criterion}(c)  a noticeably different heating response. The reason is that the density of states near the edge of the energy spectrum of a spin cluster is very low, and, as a result, the lowest 1 percent of the energy states are mostly located  relatively far from $E_g$. At the same time, the ensemble representing $T=1$ is dominated by a large contribution of only very few lowest-energy states.


\section{F. Normal heating for higher initial temperatures as a delocalization criterion for lower initial temperatures}
\label{sec:low-temperature}

In the main article, we remarked that even though our DL criterion is formulated for the infinite-temperature regime, it, possibly, constitutes not only sufficient but also necessary condition for DL to occur at finite temperatures.  Here we include additional discussion to support this conjecture.

The competition between dynamic localization and delocalization depends on the density of many-particle states associated with Hamiltonian ${\cal H}_{\rm off}$ around the energy of the initial state. The higher the initial temperature, the higher the density of states in the relevant energy range. Higher density of states favors delocalization. Lower density of states favors localization.

Let us consider a finite cluster where high-temperature initial states are delocalized under periodic driving, and then examine what happens with a low-temperature initial state, which corresponds  to an energy range where the density of states is much lower and hence the second-order perturbative effect of the periodic pulses (used to obtain the golden-rule heating rate $\Gamma$ in the supplemental section C) would not be sufficient to delocalize the system. In this case, higher-order perturbative effects of multiple kicks by ${\cal H}_{\rm on}$ are still supposed to eventually connect the states from the low-temperature range to the higher temperature range of ${\cal H}_{\rm off}$ . Therefore, as stated in the main article, we expect that ``in finite clusters with ergodic Hamiltonians,  if high-temperature states are dynamically delocalized, then,  low-temperature states are likely to ``leak'' to the high-temperature range due to higher-order effects of the perturbations by ${\cal H}_{\rm on}$...''

The above ``leakage scenaro'' is similar to Mott's scenario behind the notion of mobility edge for disordered electronic systems: Mott postulated that localized and delocalized states cannot coexist at the same energy, because localized states would eventually leak into delocalized ones [N. Mott, J. Phys. C: Solid State Phys. {\bf 20} 3075 (1987)].  We now observe that, after energy backfolding into the first Floquet zone, the low-energy eigenstates of ${\cal H}_{\rm off}$ become mixed with many more high-energy eigenstates.    The Mott postulate can then be rephrased by stating that, generically, dynamically localized Floquet eigenstates cannot coexist with dynamically delocalized ones in the same quasi-energy range.

An important outcome of our numerical simulations is that they support the ``leakage scenario'': in all examples presented in Figs.~\ref{fig:Eav}(b) and (d), the reponses of the same cluster starting from the ground state and from the thermal state with $T=1$ can be quite different, but they either both exhibit DL or both reach $T=\infty$.
The prethermalization delay mentioned in the main article can also be observed Figs.~\ref{fig:Eav}(b,d) --- for example, by comparing  the responses of the 13-spin cluster. 
One can see in this case that the heating responses starting from the ground state [Figs.~\ref{fig:Eav}(b)] and the thermal state with initial temperature $T = 1$ [Figs.~\ref{fig:Eav}(d)] reach the same asymptotic energy, but the response starting from the ground state does it noticeably slower.

We note, however, that for clusters of 10-16 spins 1/2 considered in the present work, it only takes relatively few kicks by the perturbation ${\cal H}_{\rm on}$ before the ground state and the infinite-temperature states of ${\cal H}_{\rm off}$ become connected. In principle, if one increases the size of the cluster, while keeping the perturbation strength just  slightly above the high-temperature delocalization threshold, the differences between the heating times of low-energy and high-energy states can become much larger. Whether the low-energy states can then be considered, for all practical purposes,  dynamically localized  requires further investigations.

Finally, we would like to remark that, even when the low-temperature states are not dynamically localized but exhibit the prethermalization delay with the onset of normal heating, this delay is an essentially quantum effect, which is sensitive to the size of quantum clusters. Therefore, the size of quantum clusters can, in principle, be diagnosed not only from the difference between asymptotic heating responses under weak periodic and aperiodic perturbations as proposed in the main article but also from the difference between the initial responses, which, however, requires an additional quantitative analysis extending beyond the scope of the present work.

\end{document}